\documentstyle [epsf,mnras_cite] {mn} 
\def\plotfiddle#1#2#3#4#5#6#7{\centering \leavevmode
\vbox to#2{\rule{0pt}{#2}}
\includegraphics{#1}} 
\begin{document}
\title[Spatial Distributions of Cluster Galaxies]{The Spatial and  
Kinematic Distributions of Cluster Galaxies in a $\Lambda$CDM Universe    -- 
Comparison  with Observations}
\vspace{1cm} 
\author[A. Diaferio,  et al.]
{Antonaldo Diaferio$^{2,1}$, Guinevere Kauffmann$^1$, Michael L. Balogh$^3$ 
\newauthor
Simon D.M. White$^1$, David Schade$^4$ and Erica Ellingson$^{5}$\\   
$^1$Max-Planck Institut f\"ur Astrophysik, Karl-Schwarzschild-Str. 1,
D-85741 Garching, Germany\\
$^2$ Universit\`a di Torino, Dipartimento di Fisica Generale "Amedeo Avogadro", Via P. Giuria 1,
I-10125 Torino, Italy\\   
$^3$ Department of Physics, University of Durham, South Road, Durham DH1 3LE, UK\\
$^4$ Canadian Astronomy Data Center, Hertzberg Institute of Astrophysics, 5071 W. Saanich Rd.,
Victoria, BC Canada, V8X 4M6\\
$^5$ Center for Astrophysics and Space Astronomy, University of Colorado, CO 80309, USA\\}

\maketitle
\begin {abstract}
We  combine dissipationless N-body simulations and semi-analytic
models of galaxy formation to study the spatial and kinematic distributions of
cluster galaxies in a $\Lambda$CDM cosmology.
We investigate how the star formation rates, colours and morphologies of galaxies vary as a function 
of distance from the cluster centre and compare our results with the CNOC1 survey 
of galaxies from  15 X--ray luminous clusters in the redshift range $0.18 < z < 0.55$.
In our model, gas no longer cools onto galaxies after they fall into the cluster and as a result,
their star formation rates decline on timescales of $\sim 1-2$  Gyr.
Galaxies in cluster cores have                                
lower star formation rates and redder colours than galaxies in the outer regions
because they were accreted earlier. Our colour and star formation gradients agree
with those derived from the data. The difference in velocity 
dispersions between red and blue galaxies observed in the CNOC1 clusters
is also well reproduced by the model.
We assume that the morphologies of cluster galaxies are determined solely by 
their merging histories. A merger between two equal mass galaxies produces 
a bulge and subsequent cooling of gas results in the formation of a new disk. 
Morphology gradients in clusters arise naturally, with the fraction
of bulge-dominated galaxies highest in cluster cores. The fraction of bulge-dominated galaxies
inside the virial radius depends on the mass of the cluster, but is independent
of redshift for clusters of fixed mass. Galaxy colours and star formation rates 
do not depend on cluster mass.
We compare the distributions of galaxies in our simulations
 as a function of bulge-to-disk ratio and as a function
of projected clustercentric radius  to those derived from the CNOC1 sample.      
We find excellent agreement for bulge-dominated galaxies. The simulated clusters contain too few galaxies 
of intermediate bulge-to-disk ratio, suggesting that additional processes may influence the morphological
evolution of disk-dominated galaxies in clusters.          
Although the properties of the cluster galaxies in our model agree extremely well with the data,
the same is not true of field galaxies. Both the star formation rates and the colours of
bright field galaxies appear to 
evolve much more strongly from redshift 0.2 to 0.4 in the CNOC1 field sample 
than in our simulations.

\end {abstract}

\newpage  
\section {Introduction}

In recent years, there has been a huge observational effort aimed at understanding the evolution of
galaxy populations in clusters. High resolution images from the Hubble Space Telescope
have been used to study the morphologies of cluster galaxies and to quantify 
the incidence of interacting or disturbed systems out to $z \sim 1$ (see for example
Oemler et al 1997; Lubin et al 1998) . Multi-object spectroscopy
has been used to confirm cluster membership and to study the star formation histories and
kinematics of cluster galaxies (e.g. Dressler et al 1999; Poggianti et al 1999). 
Some recent studies (e.g. Abraham et al 1996; Balogh et al 1999, Van Dokkum et al 1998)
have focused on the observed trends in star formation and morphology {\em as a function of position}
within the cluster. These studies demonstrate that there is a smooth transition from a blue, disk-dominated
population of galaxies in the outskirts of clusters to a red, bulge-dominated population
in the cluster cores.
In spite of this  wealth of new data, the physical processes responsible for driving the transformation
of galaxies in groups and  clusters remain poorly understood. Galaxy-galaxy interactions 
and mergers  (Lavery \& Henry 1988), tidal disruption (Byrd \& Valtonen 1990),
interactions with the intracluster medium (Farouki \& Shapiro 1980; Abadi, Moore \& Bower 1999) 
and repeated high-speed galaxy encounters 
with the cluster potential (``harassment''; Moore et al 1996) are all popular explanations of some
of the properties of the blue galaxies seen in intermediate redshift clusters (Butcher \& Oemler 1978).
In order to understand trends in galaxy properties as a function of environment and of  
redshift, it is necessary to consider both the physical processes that affect galaxies in dense environments
and the assembly of the cluster itself. The formation of clusters has traditionally been studied
in one of two ways: 1) using
methods based on the extended Press-Schechter (EPS) theory (Bower 1991; Bond et al 1991), 2) using
N-body simulations of gravitational clustering.    

Kauffmann (1995 a,b) used the EPS approach to show that the evolutionary history is different for
a rich cluster seen at high redshift than for a cluster of the same mass observed today. When
simplified prescriptions for gas cooling, star formation, supernova feedback and galaxy-galaxy merging were 
included in the models, the fraction of blue star-forming galaxies in rich clusters was shown to increase
with redshift. The best fit to the data was obtained for a high-density CDM cosmology. Because
clusters evolve more slowly in a low-density Universe, the  low-$\Omega$
models produced a much weaker trend in blue fraction with redshift, 
in apparent contradiction with the observations.
The disadvantage of the EPS approach is that it is not possible to model the spatial distribution of
galaxies within rich clusters. The models do not follow                         
substructure within individual dark matter halos. It is thus not possible to study whether the         
observed radial gradients in clusters arise because processes
such as ram-pressure stripping operate more efficiently in cluster cores, or because galaxies in the
central regions were accreted at an earlier epoch than galaxies on the outside.

Evrard, Silk \& Szalay (1990) modelled the spatial distribution of galaxies in
an N-body simulation of a cluster by associating galaxies with peaks in the initial density field.
They proposed that elliptical galaxies were associated with the
highest peaks and that spiral galaxies were associated with smaller peaks and demonstrated that
they could obtain a morphology-density relation in reasonable agreement with observations.
The disadvantage of this study was that galaxy formation was treated in an extremely simplistic
way and it was not possible to make close contact with observational data.

In this paper, we study the evolution of galaxies in clusters using  an N-body simulation
in which the formation and evolution of galaxies are followed using prescriptions taken directly from
semi-analytic models.
We focus on how the properties of galaxies vary as a function of
position in the clusters and how these trends evolve with redshift.
The techniques used for constructing dark matter halo merger trees from 
the simulation and the recipes used for cooling, star formation, feedback,
and galaxy-galaxy merging were described in Kauffmann et al (1999a, hereafter KCDW).
This paper also described the global properties of galaxies at $z=0$ including their luminosity functions
and two-point correlation functions.

In previous papers (KCDW; Kauffmann et al 1999b; Diaferio et al. 1999), 
we have explored two different cosmologies:
a high-density CDM model ($\tau$CDM) with $\Omega=1$, $\sigma_8=0.6$ and $H_0= 50$ km s$^{-1}$ Mpc$^{-1}$,
and a low-density model with $\Omega=0.3$, $\Lambda=0.7$ $\sigma_8=0.9$ and
$H_0 = 70$ km s$^{-1}$ Mpc$^{-1}$ ($\Lambda$CDM). 
We found that the $\tau$CDM model consistently 
failed to fit the observations as well as the $\Lambda$CDM model. In particular, $\tau$CDM produced
a field galaxy luminosity function with too many very bright galaxies (Kauffmann et al 1999a)
and was unable to reproduce the topology of the large scale galaxy distribution
(Schmalzing and Diaferio 2000).
In this paper, we only consider the more successful $\Lambda$CDM model. All quantities
are for $H_0 = 70$ km s$^{-1}$ Mpc$^{-1}$.

The simulation volume is 14000 km s$^{-1}$ on a side and contains
$\sim 140$ clusters with masses greater than $10^{14} M_{\odot}$ at $z=0$.
The clusters contain between 5000 and 80000 dark matter particles.
As discussed in KCDW,  the B-band luminosities of galaxies in the
simulation can be reliably          
determined  down to $\sim 1.5$ magnitudes below $L_*$. Because the morphologies of galaxies
depend on their detailed merging histories, galaxy types can only be determined for  
objects brighter than $\sim L_*$. To obtain reasonable statistics,                     
we stack all the clusters in the simulation and rescale the clustercentric
distances by dividing by $R_{200}$, the virial radius of the cluster. This approach follows
the one adopted by Yee et al (1996) and Balogh et al (1997,1998,1999) in a series of papers 
analyzing the observed radial trends in clusters in the CNOC1 survey.
Where explicit comparisons with observational data are made, we analyze galaxy properties
as a function of projected clustercentric distance $R_{proj}$ and exclude galaxies with
large velocity differences from the central cluster galaxy.
This will be discussed in more detail in section 4. 

In section 2, we discuss those aspects of our model that influence the evolution of cluster galaxies.
Section 3 summarizes the properties of the simulated clusters. The CNOC1 cluster sample is
discussed in section 4. Results on the cluster luminosity function, star formation gradients,
colour gradients, morphology gradients  and the kinematics of cluster galaxies are presented
in sections 5-9. Finally, we summarize our results in section 10.

\section {Model assumptions that influence the evolution of cluster galaxies}
 
In this section, we review the processes that influence the evolution of galaxies in
clusters in our model.

\begin {enumerate} 
\item {\bf Gas supply and star formation.} As discussed in KCDW, the star formation rate in a galaxy
is regulated by the rate at which gas cools from the surrounding hot halo and the rate at which supernovae
eject cold gas out of the galaxy. When a galaxy is accreted by a more massive
group or cluster, it loses its supply of infalling cold gas. Its star formation rate
then declines as its existing reservoir of cold gas is used up.
The time taken for a blue, star-forming galaxy to exhaust its fuel supply and become red, depends on the
assumed star formation timescale. In KCDW we adopted the empirically-motivated star formation law 
of Kennicutt (1998), which has  the form
$\dot{M}_* = \alpha M_{cold} / t_{dyn}$, where $M_{cold}$ is the mass of cold gas left in
the galaxy and $t_{dyn}$ is the dynamical time of the galaxy. The dynamical time of
the galaxy is defined when the galaxy was last a {\em central galaxy} in a halo, and is
given by 
$t_{dyn} =0.1 R_{200}/ V_c$,
where $R_{200}$ and $V_c$ are the virial radius and circular velocity of the surrounding halo.  
The parameter $\alpha$, which controls
the efficiency of star formation, was chosen to obtain a  cold gas mass   
of $\sim 8 \times 10^{9} M_{\odot}$ for a Milky Way--type galaxy at the present day
and has a value $\sim 0.1$.  This means that a Milky--Way  galaxy in a halo with circular
velocity $220$ km s$^{-1}$ and virial radius $0.4$ Mpc, will run out of cold gas              
$\sim 1.5$ Gyr after being accreted by a more massive halo. Note that this calculation of the 
gas consumption
timescale neglects the effects of recycling due to mass loss and supernova ejecta, which
may increase it by a factor of 1.5-3, depending upon the assumed IMF
(Kennicutt, Tamblyn \& Congdon 1994).  However, it  is also possible that ram-pressure
stripping may accelerate the rate at which
the gas is removed from galaxies (Abadi et al 1999).

As shown by Kauffmann \& Haehnelt (2000), if
$\alpha$ is a constant independent of redshift, the ratio of gas mass to stellar mass in galaxies 
evolves weakly with redshift. 
In order to reproduce the observed increase in the total
mass of cold gas in the Universe inferred from damped Lyman-alpha systems and to 
explain the strong increase in the space density of quasars and starburst galaxies
from the present day to  $z \sim 2$  Kauffmann \& Haehnelt
adopted a  redshift-dependent $\alpha$ of the form $\alpha \propto (1+z)^{-1.5}$.
In this case,  the fraction of gas converted into stars per dynamical time 
is lower for high redshift galaxies, so galaxies falling into clusters at high redshift
will take longer to exhaust their cold gas reservoirs. Somerville, Primack \& Faber (2000) showed that
the same star formation law could reproduce the observed evolution of cold gas
in damped systems and the properties of Lyman-break galaxies at $z \sim 3$.

Figure 1  compares the evolution of the gas fractions
in field and cluster  galaxies for the two star formation prescriptions.
If $\alpha$ is constant, the mean ratio of gas to stars in  galaxies more massive than
$3 \times 10^{10} M_{\odot}$ increases by a factor $\sim 2$ from $z=0$ to $z=1$.
If $\alpha \propto (1+z)^{-1.5}$, this ratio increases by a factor $\sim 10$ over the same redshift
interval. A similar effect is seen in clusters, where the gas-to-star ratios of galaxies
are typically a factor of 10 lower than in the field.

We have explored both star formation laws in this analysis. At the relatively low redshifts
of the CNOC1 cluster sample ($0.18 < z < 0.55$), there is no observationally discernable difference
in the star formation rates or colours of cluster galaxies for the two prescriptions. 
On the other hand, an evolving $\alpha$ results in slightly stronger evolution of the star formation
rates of bright field galaxies and is in  better agreement with the
CNOC1 field sample, so we adopt $\alpha \propto (1+z)^{-1.5}$ as the     
fiducial model in this paper.

\begin{figure}
\centerline{
\epsfxsize=8cm \epsfbox{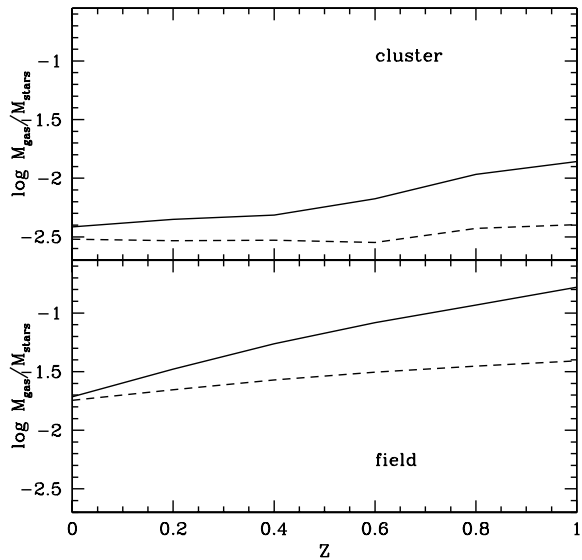}
}
\caption{\label{fig1}
\small
The evolution of the mean ratio of gas mass to stellar mass  for galaxies 
more massive than $3 \times 10^{10} M_{\odot}$
in clusters and in the field. The dashed line is for the model with constant $\alpha$ and
the solid line is for the model with $\alpha \propto (1+z)^{-1.5}$.}
\end {figure}
\normalsize

\item {\bf Definition of the cluster boundary} As described in KCDW, the clusters 
in our model are selected using a friends-of-friends group-finding algorithm 
with a linking length $b=0.2$.
This means that clusters are defined at an overdensity of $\sim 200$ times
the {\em background density} at each redshift. A galaxy is said to have been ``accreted'' by a cluster
if it is included in the list of particles that are linked together by the groupfinder.  
After a galaxy has been accreted, no further cooling of gas onto the galaxy takes place and
its star formation rate declines on a timescale that is short compared
to the Hubble time. Note that the same thing occurs for galaxies that ``escape'' temporarily
and are not included within the boundary of any halo. Balogh, Navarro \& Morris (2000) 
noted that this occurred  quite frequently in the simulations they studied. We note that in their
study, ``galaxies'' were chosen to be a random subset of the particles in the cluster
at the end of the simulation, rather than the most-bound particles of the cluster progenitors at
high redshift.

As a result of our approach, a fairly sharp transition occurs at the cluster boundary 
from a gas-rich, star-forming  population of ``field'' galaxies on the outside to a gas-poor,
red population of cluster galaxies on the inside. The cluster boundary lies  
at a  distance of $\sim 1.5-2 R_{200}$ from the cluster centre in our $\Lambda$CDM model
(Recall that $R_{200}$ is
defined relative to the {\em critical density}).
A more physically realistic treatment of the accretion process would consider how  
the hot gas surrounding infalling galaxies is shock-heated and tidally stripped from their
halos.                  

\item {\bf Merging and morphology.}
In our model, the morphologies of cluster galaxies are determined by the rate at which
they merge and this is in turn set by the merging history of the
dark matter component of the cluster. High resolution N-body simulations have shown that
if a dark matter satellite falls into a larger halo, the satellite can preserve its identity for
some time (e.g. Klypin et al 1999). Smaller satellites may survive for a long time, whereas larger 
satellites quickly sink to the centre of the halo and merge with the central mass concentration.
Since our simulations do not have sufficient resolution to follow the
orbits of individual satellites,  we have modelled the rate at which dynamical friction 
causes a satellite halo of given mass to sink to the centre of the larger halo and merge (see KCDW).
Detailed comparisons of the simple analytic formula  with N-body simulations have shown 
that {\em on average} this 
treatment provides a reasonably accurate description of  merging timescales 
(Navarro, Frenk \& White 1995; Tormen, Diaferio \& Syer 1998; van den Bosch 1999; Springel et al 2000).
However, some of the most massive satellites take significantly longer to merge than predicted
by the simple formula and this has a  significant effect on the 
mass distribution of the brightest galaxies
in the clusters (Springel et al 2000).

If two galaxies of roughly equal mass merge, the remnant is classified as a ``bulge''. The
stars in the two galaxies are combined and all the cold gas is transformed into stars in a burst
lasting $10^8$ years. Further cooling leads to the formation of a new disk component. 
We have made morphological classifications according to the B-band bulge-to-disk ratios
of the  galaxies in the simulation. Following Balogh et al (1998), we split our sample
into three classes: a bulge-dominated class (B), a disk-dominated class (D) and an
intermediate class (Int). We choose the boundaries between the classes so as to reproduce
the  fraction of {\em field galaxies} in each class in the CNOC1 data (see figure \ref{morphs1}). 
Bulge-dominated galaxies
then have $B/T > 0.4$, disk-dominated galaxies have $B/T < 0.2$ and the intermediate class
has $0.2 < B/T < 0.4$. (In the CNOC1 data, the corresponding divisions are at $B/T > 0.7$ (B),
$B/T < 0.4$ (D) and, $0.4 < B/T < 0.7$ (Int) in the $r$-band (see section 4).)  

In our models, satellite galaxies merge only with the central galaxy of the halo. We do not
consider the effect of  collisions or close  encounters between satellites.
Springel et al (2000) have found that in high resolution N-body simulations
of cluster formation,  mergers between satellites are relatively rare (only one in twenty
mergers took place between two satellite halos, rather than a satellite and the central halo). 
On the other hand, the rate of close enounters is 
high (Tormen, Diaferio \& Syer 1998; Kolatt et al 1999)
and may be sufficient to explain the observed incidence of blue galaxies with disturbed morphologies
in rich clusters at $z \sim 0.4$.  Moore at al (1999) have demonstrated that these close encounters 
may transform low surface brightness disk galaxies into dwarf spheroidals and high surface
brightness disks into S0s. 

In our models, bulge-dominated galaxies in clusters are formed by mergers occurring in smaller groups
that are later accreted by the cluster. The only galaxy affected
by mergers in the cluster itself is the central object, which grows steadily in mass by 
accreting smaller satellites. 

\item {\bf Positions of galaxies within the cluster: a reflection of incomplete violent relaxation.}
As discussed in KCDW, the galaxy formed from gas cooling in a dark matter halo is assigned
the index of the most bound particle in that halo. The galaxy is always identified with the same
particle, even after it is accreted by a larger group or cluster. 

High-resolution N-body simulations show that the dense cores of dark matter halos
often survive after they are accreted by a larger system , even after many crossing times. 
Springel et al (2000) have developed
methods of picking out surviving ``subhalos'' orbiting within a larger halo and have studied how
the subhalos are tidally stripped over time. Subhalos were studied down to a threshold of 10 particles.
In 90\% of cases, surviving subhalos still contain the most-bound particle identified before the halo 
was accreted.
This result is independent of the resolution of the simulation and it indicates that 
our technique of  using the
most-bound particle to mark the positions of galaxies within clusters is robust,
particularly for the galaxies with luminosities $ > L_*$, which always form in dark matter halos
of at least 100 particles in our simulations. 

It should be noted that radial trends in galaxy clusters {\em cannot} be studied using the simplified
procedures for assigning galaxies to halos adopted by Kauffmann, Nusser \& Steinmetz (1997) or
Benson et al (2000). These authors do not follow the merging histories of the halos in
the simulations and simply assign galaxies to a {\em random subset} of the halo particles.

\item {\bf Spatial bias and the properties of galaxies in the vicinity of clusters.}          
Clusters and rich groups correspond to high peaks in the initial field of density fluctuations 
and are consequently more highly clustered than dark matter halos of lower mass (Bardeen et al 1986).
As a result of this bias, the properties of galaxies in the vicinity of clusters will
not be the same as the properties of galaxies in the field.
Our procedure of following the formation and evolution of galaxies in the halos defined
by the dark matter simulation means that these spatial bias effects are automatically 
taken into account.

\end {enumerate}

In summary, we have made a set of extremely simple assumptions about the
gas physical processes operating  within clusters. The true situation is undoubtedly
more complicated. Our intention is to concentrate on the radial trends 
induced by the assembly of the dark matter component of the cluster. 
If these trends  disagree with the observations, we may then learn something about 
additional physical processes
at work in the cluster.

\section {The cluster sample in the simulations}
\begin{figure}
\centerline{
\epsfxsize=7cm \epsfbox{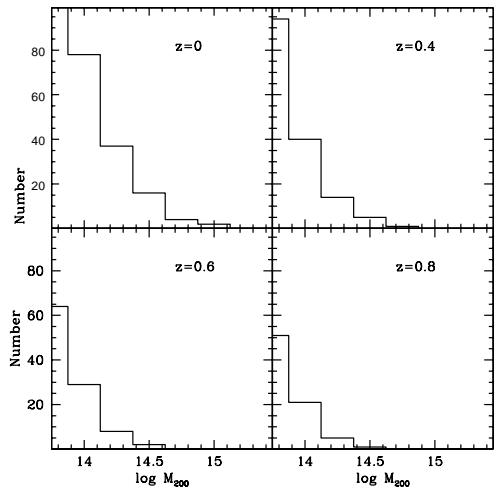}
}
\caption{\label{massfun}
\small
The number of clusters in our samples as a function of the logarithm of the virial mass 
in units of $M_{\odot}$.}
\end {figure}

We have selected dark matter halos with virial masses $M_{200}\ge 10^{14} M_{\odot}$ at
a series of different redshifts.
$M_{200}$ is the mass inside  $R_{200}$, the radius within which the mass overdensity
is 200 times the critical density. The mass distribution of the objects in our sample is
shown at a series of redshifts in Fig. \ref{massfun}.  
Although there are three clusters at z=0 with masses comparable to
the Coma cluster ($ \ge 10^{15} M_{\odot}$), by z=0.8 the most massive clusters in the simulation volume
are more comparable to Virgo. Our sample of high-redshift clusters is therefore not 
strictly comparable to samples such CNOC1, which include only 
the most X-ray luminous systems in a substantially larger volume of the Universe. 
The CNOC1 clusters range
from $3.4 \times 10^{14} M_{\odot}$ to $4.3 \times 10^{15} M_{\odot}$, with a
median mass of $1 \times 10^{15} M_{\odot}$. 
In this paper  we will attempt to indicate whether or not our
results depend on cluster selection by studying how cluster galaxy properties vary  
as a function of cluster mass in our model.

Cluster galaxies are identified in one of two ways: 1) We use the galaxy         
positions from the simulations to study the properties of galaxies as a function of physical
radius from the cluster centre. 2) We study galaxy  properties  as a function of 
two-dimensional projected radius from the central cluster galaxy. In this case, we exclude
galaxies with large velocity differences from the central cluster galaxy in exactly the same
way as in the observations (see section 4).

\normalsize

\section {The CNOC1 cluster sample}

The CNOC 1 cluster sample consists of fifteen
X--ray luminous clusters in the redshift
range $0.18<z<0.55$.  Full details of the survey are given in Yee,
Ellingson \& Carlberg (1996).

Star formation rates for the galaxies were determined from
measurements of the equivalent width of the [OII]$\lambda$3727
emission line ($W_\circ(OII)$) as described in
Balogh et al. (1998), using the calibration of Barbaro \& Poggianti
(1997). We add a small amount (0.04 h$^{-2}$ M$_{\odot}$ yr$^{-1}$) to
the measured rates to compensate for the fact that the $W_\circ(OII)$
index is slightly negative in the absence of any emission.
The statistical uncertainty in derived star formation rates
is typically $\sim$0.2 h$^{-2}$ M$_{\odot}$ yr$^{-1}$. However, there
are many systematic uncertainties in the conversion of $W_\circ(OII)$ to
a star formation rate, as discussed   
for example by Kennicutt (1998). For this reason, we choose to concentrate on 
{\em relative} star formation rates    
(for example by comparing  cluster and field galaxies),
rather than absolute ones.

Morphological parameters for the $r$--band multi-object spectrograph (MOS) images were
measured by fitting two dimensional exponential disk and
$R^{1/4}$ law profiles to the symmetrized components of the light
distribution, as described in Schade et al. (1996a, 1996b).  The
images are symmetrized to minimize the effects of nearby companions
and asymmetric structure, and a $\chi^2$ minimization procedure is
applied to the models, convolved with the image point spread function,
to obtain best fit values of the galaxy size, surface brightness and
fractional bulge luminosity (bulge--to--total, or B/T ratio).
Simulations show that the B/T measurements are reliable to within about
20\% for images of this quality (Schade et al 1996a,b).   
These measurements are the same as used in Balogh et al (1998).

A proper consideration of selection effects is important, because
spectra are not
obtained for all the galaxies that were observed photometrically.  We correct
for these effects using the statistical weights discussed in detail in
Yee et al. (1996).  The main selection criterion is apparent
magnitude; a smaller fraction of faint galaxies are observed
spectroscopically, relative to brighter galaxies.  The magnitude
weight $W_m$ compensates for this effect.  A second order, geometric
weight $W_{xy}$ is computed which compensates for effects such as the
undersampling of denser regions, and vignetting near the corners of
the chip.  Finally, a color weight $W_c$ is computed to account for
the fact that bluer galaxies are more likely to show emission lines, thus
facilitating redshift determination. This is a small effect
and we have checked that the inclusion of this weight does not
significantly affect any of the results discussed in the present
paper.  Both $W_{xy}$ and $W_{c}$ are normalized so that their mean is
1.0 for the full sample.  In all of our analyses, each
galaxy in our sample is weighted by $W_{\rm spec}=W_m\times
W_{xy}\times W_c$ to correct for these selection
effects in a statistical manner.

Gunn $g$ and $r$ photometry from the MOS images are available;
the  photometric uncertainties in these measurements at the
spectroscopic  limit are about $\sim 0.1$ mag.
Absolute $r$ magnitudes ($M_r$) are calculated from the photometry
for our adopted  cosmology ($\Omega=0.3$, $\Lambda=0.7$)  
and k--corrections are made based on the
{\em g-r} colors and the model spectral energy distributions of
Coleman, Wu \& Weedman (1980), convolved with the filter response
function, for four, non--evolving spectral types (E/S0, Sbc, Scd and
Im).   We chose an absolute magnitude limit of $r=-19.3+5 \log{h}$,
which  corresponds to $M_R=-20.5$ for $h=0.7$, assuming $r-M_R\approx
0.45$ (Fukugita et al. 1995). This magnitude limit is chosen 
because it correponds to the faintest galaxies for which we can                                 
accurately model both star formation rates and morphologies in the simulation.
Rest frame $(g-r)_{\circ}$ colours are
computed from  the colour--redshift relations in Patton et al. (1997),
which are fits to the  colour k--corrections of Yee et al. (1996). 
The corresponding rest frame $(B-V)_{\circ}$ colour is derived  by
linearly interpolating the published values in Fukugita et al. (1995).

Cluster members are considered to be those galaxies with velocity
differences from the brightest cluster galaxy that are 
less than 3$\sigma(r)$, where $\sigma(r)$ is the cluster
velocity dispersion as a function of projected radius $r$
determined from the mass models of Carlberg, Yee \& Ellingson (1997), which
are based on the Hernquist (1990) model.
Field galaxies are selected to be those with velocities greater than
6$\sigma(r)$.  Our final sample (excluding the central galaxies of
each cluster and those few galaxies for which good fits to the light
profile could not be found) consists of 557 cluster galaxies and 344
field galaxies.

Clustercentric distances are normalised to $R_{200}$, allowing us
to combine all 15 clusters in one sample, as in Balogh et al. (1998,1999).  
Since most of the CNOC1 clusters are well sampled within $R_{200}$,
asphericities and substructures within individual
clusters are averaged out when the full sample is stacked and renormalized
in this way (Yee et al. 1996).  However, this is not true beyond $R_{200}$, where data
was obtained for only 7 clusters.  Furthermore, most (78\%) of this
data comes from only 3 clusters: Abell 2390
and MS1231 at $z<0.3$, and MS1512 at $z>0.3$.  Therefore,
care must be taken in interpreting the data beyond $R_{200}$, where it
may not be considered  to be a fair statistical average of the CNOC1 sample.

\section {Luminosity evolution}

\begin{figure}
\centerline{
\epsfxsize=10cm \epsfbox{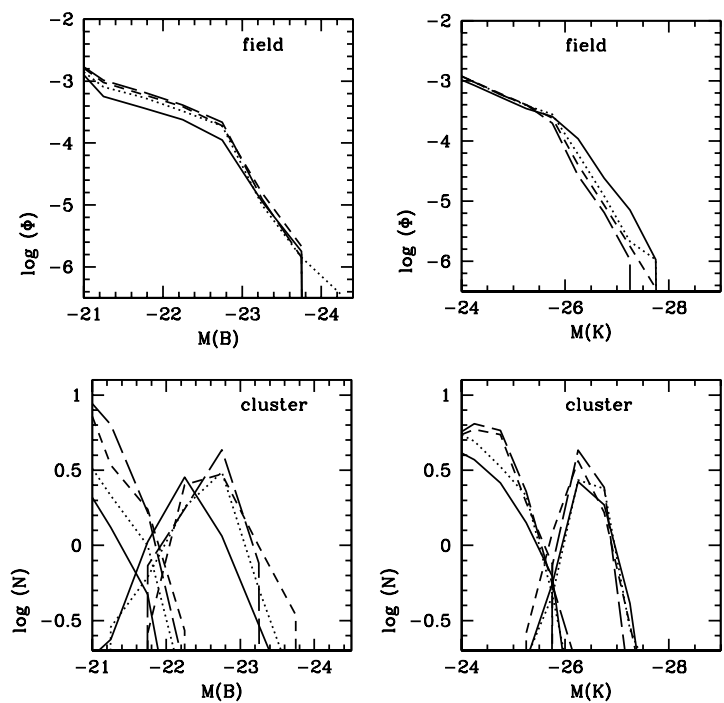}
}
\caption{\label{lumfun}
\small
{\em Top:} The evolution of the B- and K-band luminosity functions of field galaxies. The number of galaxies
per comoving Mpc$^3$ per 0.5 magnitude interval is plotted at z= 0, 0.4, 0.6 and 0.8 (solid, dotted,
short-dashed, and long-dashed curves).
{\em Bottom:} The evolution of the B- and K-band luminosity functions of cluster galaxies.
Results are shown separately for central galaxies (the bright bumps) and for the rest
of the cluster galaxy population. Results
are shown at the same redshifts. 
We plot the number of galaxies inside $R_{200}$ per 0.5 magnitude interval, scaled to a cluster 
of $10^{15} M_{\odot}$.} 
\end {figure}

In figure \ref{lumfun}, we compare the evolution of the luminosities of field and cluster galaxies        
in the simulations in the rest-frame B and K-bands.
The  B-band luminosities of galaxies are
sensitive to their star formation rates,  whereas their K-band
luminosities provide a better measure of their stellar masses (Kauffmann \& Charlot 1998). 
The field galaxy luminosity function is calculated using all galaxies in the simulation
volume. Only galaxies at physical distances less than $R_{200}$ from the centres of clusters more
massive than $10^{14} M_{\odot}$ are used in the computation of the cluster 
luminosity functions. 
In order to account for the fact that the mass distribution of clusters in the simulation
changes with redshift, 
we weight each cluster galaxy by $10^{15} M_{\odot}/ M_{200}$,
where $M_{200}$ is the virial mass of its parent cluster. In other words,  
we plot the evolution of the number
of galaxies in a given magnitude interval per $10^{15} M_{\odot}$ of cluster mass (assuming
that the number of galaxies in the cluster scales roughly in proportion to its mass
(see Seljak 2000, Sheth \& Diaferio 2000)).
For the field sample, we plot the evolution of the number of galaxies in a given magnitude
interval per unit comoving volume. Note that we only plot the luminosity functions
at bright magnitudes ($L > L_*$) where our simulation is complete
{\em at all redshifts}.

Recently, Springel et al (2000) have used a very high resolution simulation of the formation
of a single cluster to show that analytic estimates  of merging timescales of the kind used 
in this paper often underestimate the time taken    
for galaxies of near-equal mass to merge and that this can produce an overly massive
central cluster galaxy. When the  merging of satellites was followed explicitly in the simulation,
the central galaxy was $\sim 1$ mag fainter than when they used the recipes employed here.
The shape of the luminosity function was also a
much better fit to a Schechter function.
Work is currently in progress to obtain an improved parametrization of the merging
timescales of satellites using these simulations (Springel, in preparation).  
In figure \ref{lumfun}, we have plotted separate luminosity functions for central cluster
galaxies (note that objects in clusters 
less massive  than $10^{15} M_{\odot}$ are weighted by factors larger than 1), 
and for the rest of the cluster population. 

We find that the B-band luminosities of bright cluster galaxies undergo stronger luminosity evolution
than those of field galaxies. The number of bright(star-forming) galaxies in clusters increases
at high redshift when viewed in the B-band. 
In the rest-frame K-band, the luminosity function of cluster galaxies evolves very little.
We find that massive clusters contain massive galaxies, even at high redshifts.
The decrease in the space density of massive galaxies in the field at high redshifts
simply  reflects the  
decrease in the {\em global space density} of clusters in the simulation. 
We caution that most of the apparent evolution in the field over this redshift range occurs
at the very brightest magnitudes corresponding to those of central cluster galaxies. As we have discussed,
the predicted magnitudes of these galaxies are not secure.                 
However, our results agree qualitatively  with a
recent analysis  by De Propris et al. (1999), who find no significant evolution of 
the numbers of bright K-selected
galaxies in clusters out to $z \sim 1$.

\begin{figure}
\centerline{
\epsfxsize=7cm \epsfbox{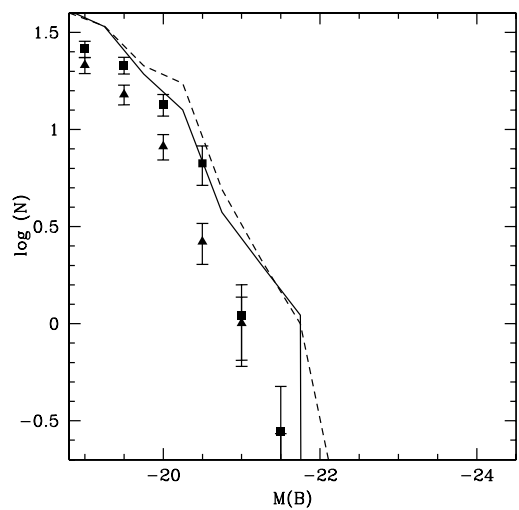}
}
\caption{\label{lum}
\small
Comparison of the B-band cluster luminosity functions at $z=0.2$ (solid line) and $z=0.4$ (dashed line)
in the simulations, with the luminosity functions of CNOC1 clusters at $z<0.3$ (squares)
and $z>0.3$ (triangles).  
We plot the number of galaxies inside $R_{200}$ per 0.5 magnitude interval, scaled to a cluster 
of $10^{15} M_{\odot}$.} 
\end {figure}

In figure \ref{lum}, we compare the luminosity functions of the simulated clusters
with those of the  CNOC1 clusters.
To compute the observed luminosity functions, we weight each galaxy by $W_{\rm spec} \times W_{\rm mass}$,
where $W_{\rm spec}$ is defined in Section 4, and $W_{\rm mass}$ is the inverse of the
virial mass of the cluster, in units of 10$^{15}$ $M_\odot$, taken from Carlberg et al. (1996).  These
weights are summed in 0.5 magnitude bins, for all galaxies within $R_{200}$,
and divided by the number of clusters in the sample (7 at $0.18<z<0.3$ and 8 at $0.3<z<0.6$).
Uncertainties are computed assuming Poisson uncertainties on the unweighted numbers.
Central cluster galaxies have been excluded in the computation of the luminosity functions
in the simulations and in the data.
A more detailed discussion of the luminosity function of cluster galaxies in the CNOC1 survey
will be given in Yee et al (in preparation).

The shape of the luminosity function in the model agrees quite well with the observations,
but its amplitude is a factor $\sim 1.5$ too large.
Given the uncertainties in deriving the masses of the observed clusters and the fact that      
the clusters in the simulations have smaller masses than in the data, the agreement
obtained is remarkably good.

\normalsize

\normalsize

\section {Star formation gradients}

\subsection {Gradients in Three Dimensions}
\begin{figure}
\centerline{
\epsfxsize=8cm \epsfbox{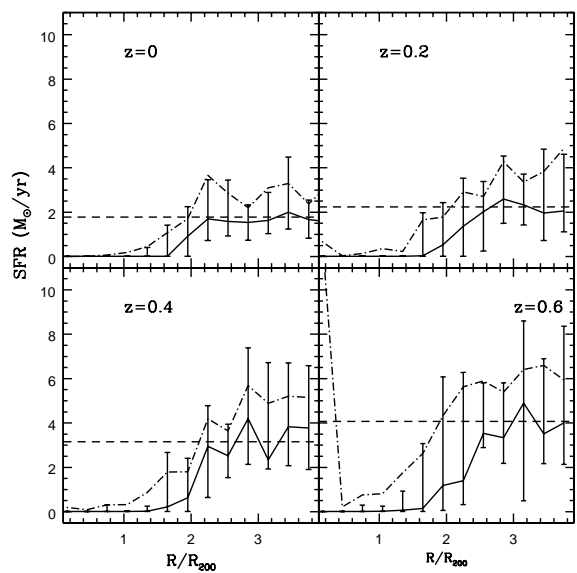}
}
\caption{\label{sfrgrad}
\small
The median star formation rate for galaxies with R-band magnitudes
less than -20.5  is plotted as a function
of $R/ R_{200}$, where $R$ is the physical distance from the cluster centre.
The error bars show the 25th to 75th percentiles of the distribution.
The straight dashed line shows the median star formation rate of galaxies in the field.
The dot-dashed line shows the {\em mean} star formation rate as a function of $R_{200}$.}

\end {figure}

The solid curve in figure \ref{sfrgrad} shows the median star formation rate of 
galaxies with $M_R < -20.5$                     
as a function of their physical  distance from the central cluster galaxy.                      
Error bars show the 25th to 75th percentiles of the
distribution and the straight dashed line shows the median star formation rate of field galaxies
at the same redshift for comparison. 
Galaxies have very widely varying star formation rates. As a result,  
the median tends to be more stable than the mean (shown as a dot-dashed line on the plot), 
which is often dominated by a few
starbursting systems. 

There is a strong trend in star formation rate  with radius. Galaxies in the centres of clusters have little or
no ongoing  star formation. Both the median and mean star formation rates of cluster
galaxies increase with radius, reaching the field value
at a distance of 2-3 $R_{200}$ from the cluster centre.  
The number of strongly star-forming galaxies at the centres of clusters increases with redshift. 
In figure \ref{sfrgrad}, this is seen as an increase in the mean star formation rate in  
cluster centres from the present day to $z \sim 0.6$.
             
\begin{figure}
\centerline{
\epsfxsize=8cm \epsfbox{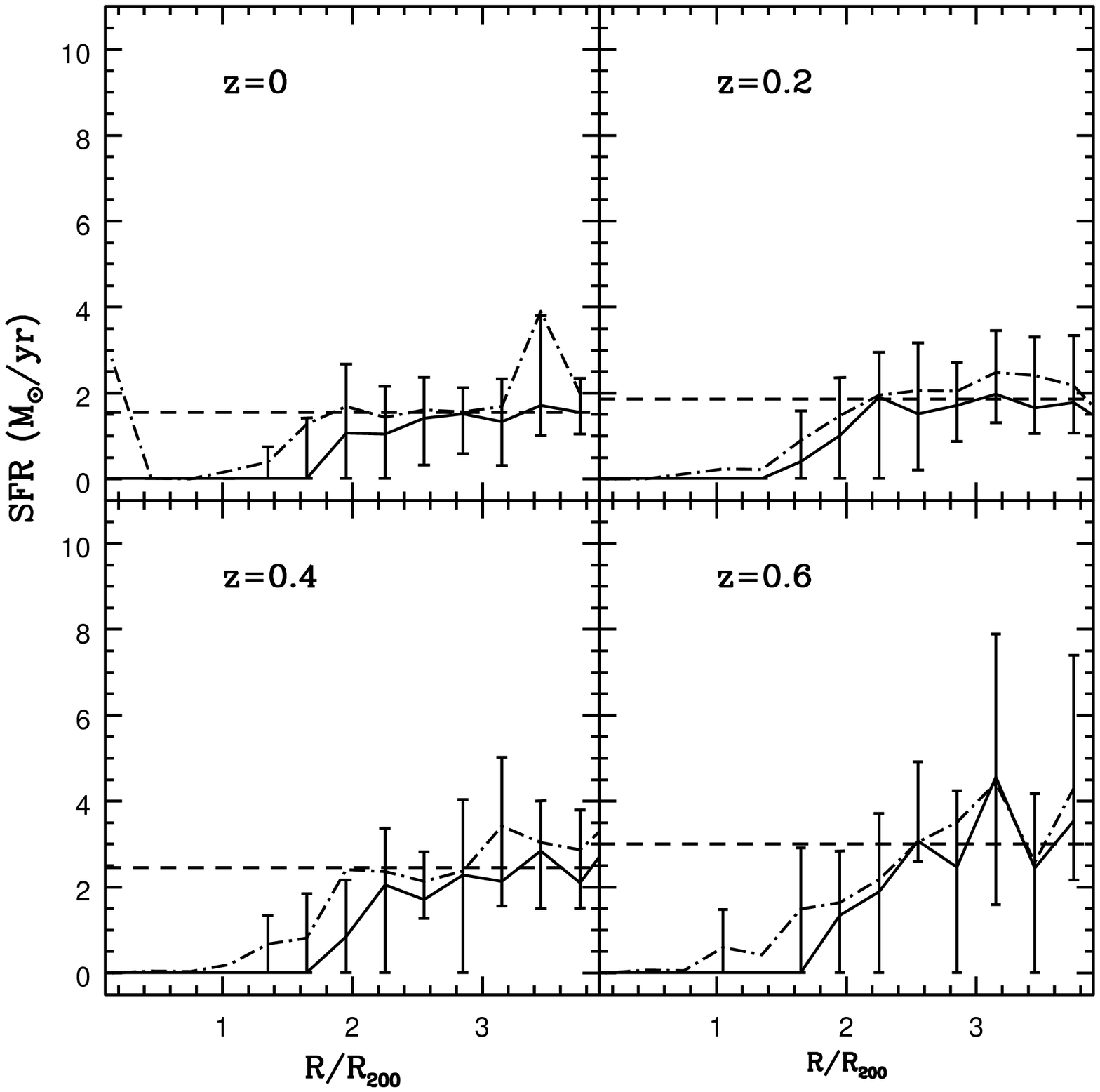}
}
\caption{\label{sfrgrado}
\small
As in figure \ref{sfrgrad}, but for a model
in which  $\alpha$ is constant .}
\end {figure}

The primary reason for this lies in our chosen parametrization of star formation: 
$\dot{M}_* = \alpha (1+z)^{-3/2} M_{cold} / t_{dyn}$. As  discussed in section 2,
galaxies at high redshift contain more gas and take longer to run out of fuel 
once they fall into a cluster. In contrast, figure \ref{sfrgrado} shows the star formation gradients at
the same redshifts if $\alpha$ is held constant. In this case, the effect is weaker.

\normalsize

\normalsize

We conclude that the hierarchical assembly of clusters in a $\Lambda$CDM cosmology can drive 
strong radial gradients in the 
star formation rates of cluster galaxies. Evolution in the number of star-forming galaxies
at the centres of clusters  at redshifts below 1 would   
indicate that galaxies are either more gas rich at high redshift or that processes 
that strip gas and quench star formation operate less efficiently in high redshift clusters.

\subsection {Comparison with Data}
Balogh et al (1998) have analyzed  radial trends in the [OII] equivalent widths of  galaxies
in the CNOC1 sample.
They find that the mean rest-frame equivalent
width of the [OII]$\lambda$ 3727 emission line decreases by more than a factor of 10
from the outskirts of the clusters to the innermost regions. Even at distances of 1-2 virial radii
from the cluster centres, galaxies have lower mean [OII] equivalent width than in the field.
Ellingson et al (1999) have performed a principal component analysis of the CNOC1 spectra
and also find strong gradients in the stellar populations of cluster galaxies.

\begin{figure}
\centerline{
\epsfxsize=8cm \epsfbox{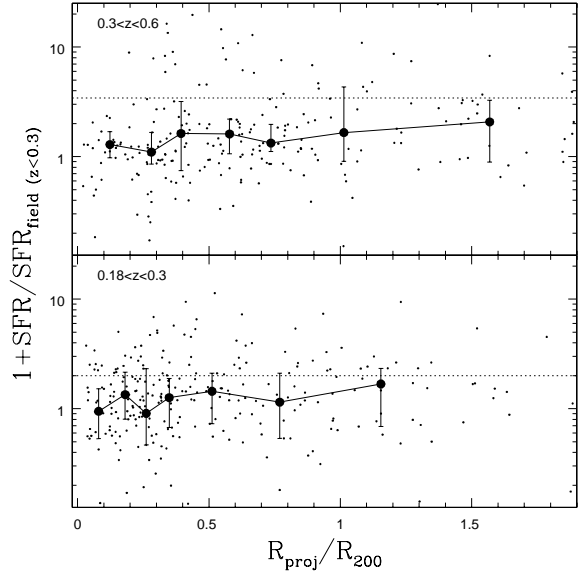}
}
\caption{\label{sfr_norm}
\small
The star formation rates of galaxies with $M_R < -20.5$ are plotted as a function of
scaled projected radius $R_{proj}/R_{200}$ for the CNOC1 sample. 
The star formation rates have been scaled by dividing
by the median star formation rate of field galaxies of the same magnitude in the 
{\em low redshift sample} and adding 1 (so that negative star formation rates do not
look pathological in the plot). The filled circles show the median star formation
rate as a function of $R_{proj}$ and the error bars show the 25th to 75th percentiles of
the distribution. The dotted line shows the median star formation rate of field galaxies
with $M_R < -20.5$ at the same redshift.}
\end {figure}

\begin{figure}
\centerline{
\epsfxsize=8cm \epsfbox{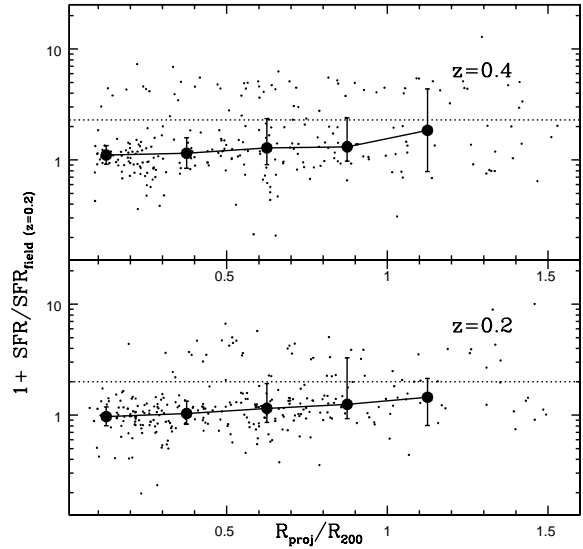}
}
\caption{\label{baltest1}
\small
The same as figure \ref{sfr_norm}, except for the simulated clusters.}
\end {figure}

In figures \ref{sfr_norm} and \ref{baltest1} we compare the star formation gradients as a function
of {\em projected radius}  obtained 
for the CNOC1 and simulated clusters. We have used the same star formation
prescription as in figure \ref{sfrgrad} and have selected only
galaxies with  R-magnitudes brighter than -20.5.
We also add a 1$\sigma$  error of 0.2 h$^{-2}$ M$_{\odot}$ yr$^{-1}$ to the star formation rates of the
simulated galaxies. One consequence is that the differences seen                                
in figures \ref{sfrgrad} and \ref{sfrgrado} are no longer observable and the
two different star formation prescriptions give almost indistinguishable
results at $z< 0.6$. We scale the star formation rates of our cluster  galaxies                    
by dividing them by the median star formation rate of field galaxies
in the low redshift sample. 
Finally, we randomly select the same number of galaxies in each radial bin as in the data.

The star formation rate distributions of cluster galaxies relative to the field look quite
similar in the simulations and in the data. The median star formation rate is very close
to zero near the cluster centre and rises to about one half the field value at
$R_{proj} \sim R_{200}$. This is true in both the low-- and high-redshift clusters.
It should be noted that the trend in star formation rate as a function of projected
clustercentric radius is much less abrupt than the trend as a function of physical
radius from the cluster centre (Fig. \ref{sfrgrad}). This means that it is    
important to include projection effects when carrying out detailed
comparisons between simulations and data.

\normalsize

\normalsize

\begin{figure}
\centerline{
\epsfxsize=6cm \epsfbox{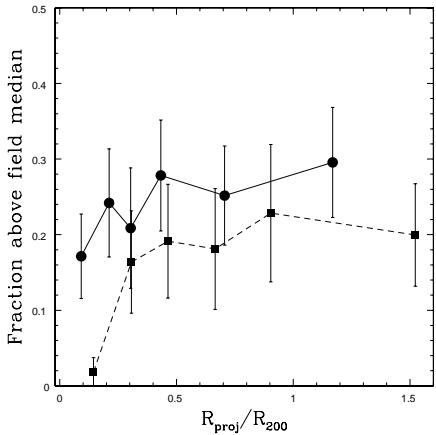}
}
\caption{\label{fractions}
\small
The fraction of galaxies in CNOC1 clusters  with $M_R < -20.5$ and star 
formation rate greater than the field median (at the same redshift)
is plotted as a function of scaled projected radius $R_{proj}/R_{200}$. 
Circles show results for clusters at $0.18 < z < 0.3$ and squares are for
clusters with $0.3 < z < 0.55$. Error bars are computed from jackknife estimation.}
\end {figure}

\begin{figure}
\centerline{
\epsfxsize=6cm \epsfbox{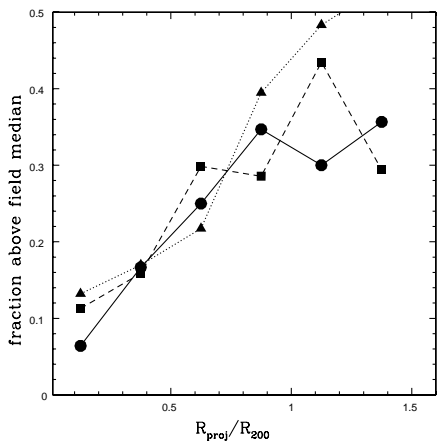}
}
\caption{\label{frac}
\small
The same as figure \ref{fractions} except for the simulated clusters. Circles are for clusters
at $z=0.2$, squares for clusters at $z=0.4$ and triangles for clusters at $z=0.8$.}

\normalsize
\end {figure}

One striking difference between the simulations and the data is that the median star
formation rate in the {\em field} appears to increase with redshift much more strongly in the data 
than in the model.
This is apparent in figure \ref{sfr_norm}, where the median star formation rate increases by a factor
of $\sim 3$ over the redshift range considered, while there is very little increase apparent
in figure \ref{baltest1}.   
In figures \ref{fractions} and \ref{frac}, we plot the fraction of cluster galaxies with star formation
rates greater than the median rate in the field  
as a function of projected clustercentric radius.
At $z \sim 0.2$, the simulation results agree with the observations quite well. The fraction
of galaxies with star formation greater than the field median increases from $\sim 0.1$ at the
cluster centres to $\sim 0.3$ at $R_{proj} = R_{200}$. At $z \sim 0.4$, the fraction of galaxies
in the CNOC1 clusters with star formation rates greater than the field median has {\em decreased}.       
This comes about because the star formation rates of galaxies in clusters do not appear
to evolve significantly over this redshift interval (figure \ref{sfr_norm}), but the median
star formation rate in the field has gone up by more than a factor of two. We do not see the same      
effect in the simulations. When scaled to the field, cluster star formation rates          
evolve very little.  At  $z =0.8$, there is a hint that the number of strongly star-forming
galaxies at the outskirts of the clusters has begun to increase.

We caution that the field samples in the CNOC1 survey are small, particularly
at $z < 0.3$. It will be interesting to see whether the strong  evolution
in the CNOC1 field  sample is confirmed by the CNOC2 survey (Lin et al 1999).

\normalsize

\section {Colours and Mass-to-light Ratios}

\subsection {Colour Gradients}

It has been known for some time that galaxies in the inner regions of clusters are redder
than galaxies in the outer regions (Butcher \& Oemler 1984; Millington \& Peach 1990).
Recently, large redshift surveys such as CNOC1 have reduced uncertainties in colour    
measurements due to the subtraction of background galaxies
and have enabled the colour gradients of clusters to
be studied in detail out to larger radii (Abraham et al 1996; Carlberg, Yee \& Ellingson 1997).

\begin{figure}
\centerline{
\epsfxsize=8cm \epsfbox{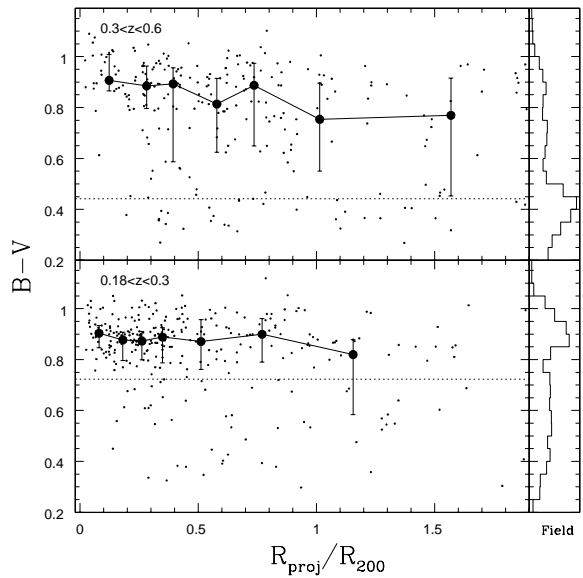}
}
\caption{\label{BV}
\small
The  rest-frame $B-V$ colours of CNOC1 galaxies with $M_R < -20.5$                                  
as a function of $R _{proj}/R_{200}$. Filled circles show the median colour
as a function of $R_{proj}$ and  error bars indicate the 25th to 75th percentiles of the distribution.
The dotted line is the median colour of field galaxies at the same redshift.}
\end {figure}

\begin{figure}
\centerline{
\epsfxsize=8cm \epsfbox{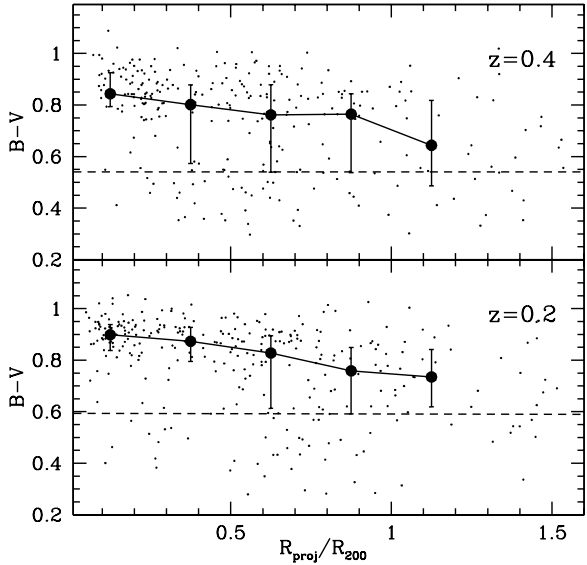}
}
\caption{\label{balcol}
\small
As in figure \ref{BV}, except for galaxies in the simulated clusters.}
\end {figure}

In figures \ref{BV} and \ref{balcol} we compare the rest-frame $B-V$ colours of cluster
galaxies as a function of $R_{proj}/R_{200}$.
Once again we have selected galaxies with rest-frame R-band magnitudes brighter
than -20.5. We have added a 1$\sigma$ error of 0.07 mag in $B-V$ to the colours
of the simulation galaxies in order to mimic the photometric uncertainties in the data.
It is reassuring that our comparison of the colour gradients leads to exactly 
the same conclusion as the comparison of the star formation gradients. 
The colours of cluster galaxies in the simulation agree quite well
with the data,
but there is much stronger evolution in the colours of field galaxies in the CNOC1 survey.
At $z=0.2$, field galaxies in the simulation are
0.1 mag too blue and at $z=0.4$ they are 0.1 mag too {\em red}.
We note that the model colours do not include any correction for dust extinction.

\subsection{A comment on the Butcher-Oemler Effect}

A major motivation for studying galaxy populations in clusters at 
intermediate redshifts has been the so-called Butcher-Oemler effect.
Butcher \& Oemler (1984) studied the fraction of bright galaxies
$f_B$ contained in the inner regions of clusters that had rest-frame $B-V$ colours at least
0.2 mag bluer than the locus of the red elliptical galaxy population. They found a strong increase
in $f_B$ from values of only a few percent in 
redshift zero clusters to $f_B \sim 0.25$ in clusters at $z=0.4$. This rapid evolution in
the galaxy populations in clusters is puzzling, especially in light of other 
observational evidence indicating that the abundance of massive clusters has
undergone only mild evolution since  $z \sim 0.5$ (e.g. Bahcall, Fan \& Cen 1997).
Kauffmann (1995b) showed that a strong increase in $f_B$ over this redshift
range  is expected only in cosmologies where clusters assemble relatively late,
such as a high-density ($\Omega_{matter}=1$) CDM cosmology with low normalization
($\sigma_8 =0.4$), or a mixed dark matter (MDM) cosmology. In low-density
CDM cosmologies, rather little evolution in $f_B$ was obtained out to $z=0.4$.

Figure \ref{BV} demonstrates that the Butcher-Oemler effect {\em is observed}
in the CNOC1 cluster sample (see also Ellingson et al 2000; Yee et al, in preparation). 
Although the median colour of cluster galaxies
differs very little between the high- and low-redshift samples, the 25th percentile  
of the colour distribution is considerably bluer for clusters with $0.3 < z < 0.55$ 
than it is for clusters with $0.18 < z < 0.3$. A similar effect is seen in the
simulated clusters, but it is significantly weaker. It should  be noted
that the colour distribution of CNOC1 {\em field galaxies} has also shifted significantly bluewards
over this redshift range.  

\begin{figure}
\centerline{
\epsfxsize=8cm \epsfbox{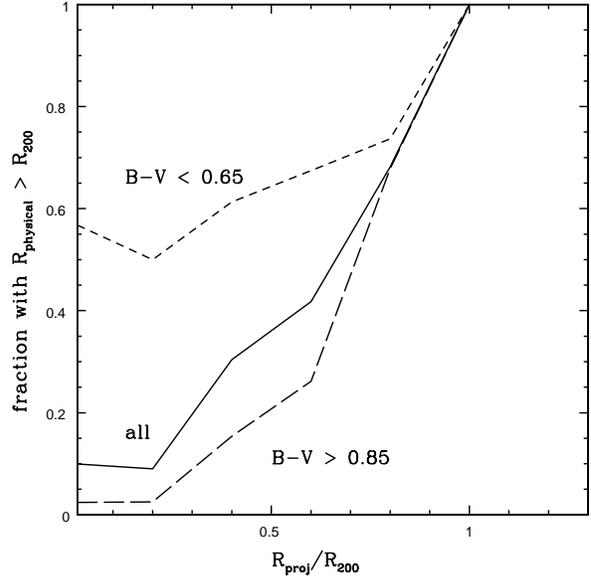}
}
\caption{\label{vel1}
\small
The  fraction of simulated  galaxies that lie at physical distances larger than $R_{200}$ from
the cluster centre as a function of scaled projected radius $R_{proj}/R_{200}$. 
Galaxies are selected to  have R-band magnitudes brighter than -20.5.
The solid line shows results for all galaxies, the long-dashed line for galaxies with $B-V > 0.85$
and the short-dashed line for galaxies with $B-V < 0.65$.}
\end {figure}

As discussed previously, galaxies with large velocity differences from  central cluster galaxies
have been  excluded from the cluster sample.
Some fraction of the galaxies will nevertheless be
interlopers from the field.  We can study this in detail in our simulated cluster sample,
where we have full position and velocity information for all galaxies, but we
have selected cluster members using the same procedures as in the observations.
In figure \ref{vel1}, we plot the fraction of galaxies that lie at  physical
distances larger than $R_{200}$ as a function of projected clustercentric distance.
This fraction goes to 1 at $R_{proj} = R_{200}$. At the centres of clusters, the
interloper fraction in the total sample is small ($\sim$ 10\%). The fraction
of red ($B-V > 0.85$) cluster members that are interlopers is even smaller ( $\sim$ 3 \%).
However, more than 50\% of galaxies  with $B-V$ colours 0.2
magnitudes bluer than the locus of bulge-dominated galaxies are interlopers.                                    

We conclude from this analysis that caution must be exercised in interpreting the physical
origin of the Butcher-Oemler effect. At least part of the
effect may be intrinsic to the field rather than to the cluster environment, and projection
effects must be taken into account.

\subsection {The Colours of Bulge-- and Disk--Dominated Galaxies as a Function of Clustercentric Radius }

Early-type galaxies in nearby rich clusters follow a tight colour-magnitude relation (Bower,
Lucey \& Ellis 1992), Mg$_2$--$\sigma$ relation (Guzman et al 1992) and Fundamental Plane
(eg. Jorgensen, Franx \& Kjaergaard 1993). The small intrinsic scatter in these relations
have been used as an argument that the early-type galaxies in these clusters formed
at high redshift, or that their formation was synchronized (Bower, Lucey \& Ellis 1992).

Kauffmann \& Charlot (1998) have demonstrated that models in which early-type galaxies
form by merging can also reproduce the observed scatter  in the
colour-magnitude relation of cluster ellipticals. 
One prediction of their models was that field ellipticals
should form later than cluster ellipticals and should thus have younger stellar
populations and exhibit greater scatter in colours and line indices.

\begin{figure}
\centerline{
\epsfxsize=8cm \epsfbox{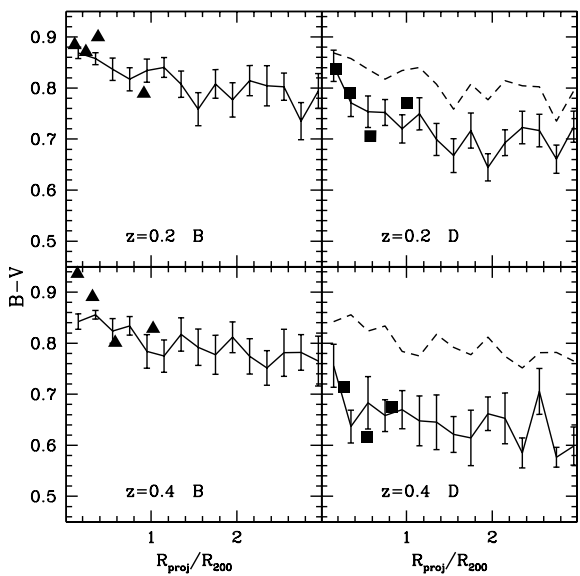}
}
\caption{\label{scatellip}
\small
Average rest-frame $B-V$ colour as a function of projected
clustercentric radius for B galaxies ($B/T > 0.4$)
and D ($B/T < 0.2$) galaxies with R-band magnitudes brighter than -20.5 in the simulations.
The error bars show the intrinsic  1$\sigma$ scatter in colour in each radial bin. The dashed lines
in the right-hand panels are simply repeats of the solid lines in the left-hand
panel, so that the reader is able to see the offset between the colours of the two classes.
The triangles and squares show average colours of B and D galaxies in the CNOC1 clusters.}
\end {figure}

There is some observational evidence in favour of younger ages for field ellipticals 
(Menanteau et al 1999; Schade et al 1999; Kodama, Bower \& Bell 1999; Trager 1999), 
although some studies yield conflicting
results (see for example Bernardi et al 1998 ; Kochanek et al 1999).     
Part of the reason for the discrepancy between different studies may lie in differing definitions     
of ``field'' versus ``cluster''.
This ambiguity  may be overcome
by studying trends in the colours of ellipticals as a function of distance from
the cluster centre. Our model predictions for the colours of bulge-dominated (B) and  
disk-dominated (D) galaxies
as a function of $R_{proj}/R_{200}$  are shown in figure \ref{scatellip}. We have again
selected galaxies with R-band magnitudes brighter than -20.5. 
In this plot, we have not added any artificial photometric errors, so the error bars represent the
{\em intrinsic} 1$\sigma$ scatter in colour of the objects in each radial bin.
We  have also plotted the average colours of B and D galaxies in the CNOC1 clusters for comparison.
The colours of both types of galaxies become bluer further from the cluster centre, but
B galaxies are on average $\sim 0.1$ mag redder than D galaxies at all radii. 
The colour difference between
the two classes appears to increase in high redshift clusters. This is also seen in
the observations.
Near the centres of clusters, B galaxies exhibit considerably smaller scatter in colour
than D galaxies, but in the outer regions both types have similar scatter.              
Because the intrinsic scatter in  colour predicted by our models  
is small compared with the photometric uncertainties of the
CNOC1 data, we do not show the scatter in the observed colours in figure \ref{scatellip}.

\subsection {Colour as a Function of Cluster Mass}

It is interesting to investigate whether the mean colours of galaxies depend on the mass of
the cluster in which they are located. Naively, one would expect  galaxies in more
massive clusters to be older and hence redder, simply because galaxies form 
at higher redshifts if they are embedded in a more 
overdense region of the Universe (Bardeen et al 1986). 

\begin{figure}
\centerline{
\epsfxsize=8cm \epsfbox{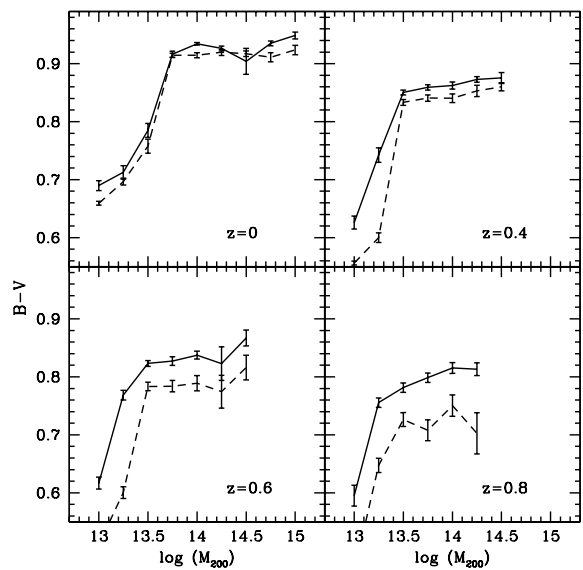}
}
\caption{\label{colmass}
\small
Average rest-frame 
$B-V$ colour of bulge-dominated (B) (solid line) and disk-dominated (D) (dashed line) 
galaxies as a 
function of the mass of the halo in which they are located. Error bars show the $1-\sigma$
scatter in the intrinsic colour of the galaxy.    
Only galaxies with $M_R < -20.5$ that lie
inside $R_{200}$ are included in the samples.}
\end {figure}

Our model predictions are shown in figure \ref{colmass}. We plot the mean rest-frame B-V colour
of bulge-dominated ($B/T > 0.4$) and disk-dominated ($B/T < 0.2$) galaxies as a function
of the mass of the halo in which they are located. We select only galaxies with
$ M_R < -20.5$ located 
less than $R_{200}$ from the cluster centre. There is a strong dependence of colour
on halo mass for halos less massive than $5 \times 10^{13} M_{\odot}$.
This is because the galaxies in the sample undergo a transition from
``central'' galaxies that accrete gas from the surrounding halo and have ongoing star
formation, to  ``satellite'' galaxies that have been stripped of their
hot gas halos and have no ongoing star formation. The precise halo mass  at which
this transition occurs thus will depend sensitively on the masses of the
galaxies in the sample. For halos more massive than $5 \times 10^{13} M_{\odot}$,
the dependence of galaxy colour on cluster mass is relatively weak. 

We note that the colours of disk-dominated  galaxies as a function of cluster mass 
may place important constraints on
gas removal in galaxies by ram-pressure stripping  because the effect
depends strongly on the mass of the cluster.

\subsection {Mass-to-light Ratios}

\begin{figure}
\centerline{
\epsfxsize=7cm \epsfbox{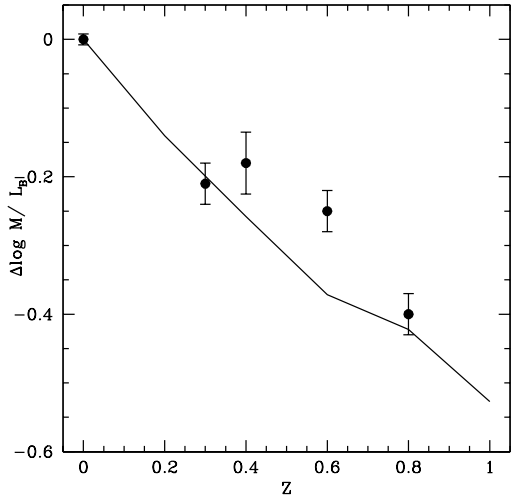}
}
\caption{\label{mlight}
The evolution of the B-band mass-to-light ratio of bulge-dominated ($B/T > 0.4$) galaxies in
clusters. The solid line shows the shift in the mean M/L in the simulated galaxies.
The points with error bars are taken from Van Dokkum et al. (1998b)}
\small
\end {figure}

The evolution of the stellar mass-to-light ratios of early-type galaxies provide a more sensitive test of the ages
of their stellar populations than their colours, mainly because the mean luminosity
evolution of an old stellar population  is large compared to its mean colour evolution.
Van Dokkum et al (1998b) find slow luminosity evolution of early-type galaxies to $z=0.8$
and show that if one were  comparing similar kinds of objects at all redshifts,
one would infer that the majority of their stars were formed at $z > 1.7$ for a Salpeter IMF and
a cosmology with $\Omega=0.3$ and $\Lambda=0.7$. 
In figure \ref{mlight}, we compare the evolution of the B-band
M/L ratios of bulge-dominated galaxies in our simulated 
clusters with data from Van Dokkum et al (1998b) plotted for 
a $\Lambda$CDM cosmology.  We have selected galaxies with rest-frame B-band magnitudes less than
than -21.5 (corresponding to galaxies with $I < 22.1$ in the observed sample)  
 and $B/T > 0.4$ that lie at distances less than
$R_{200}$ from the cluster centre.  As can be seen, the agreement is good.

\normalsize

\section {Morphologies and Morphology Gradients}\label{sect.morph}

\subsection {Evolution of Morphology Gradients}
In nearby clusters the morphologies of galaxies  are correlated with
distance from the cluster centre. Whitmore, Gilmore \& Jones (1993) have studied the
relation between morphology and clustercentric radius in Dressler's sample of 6000 galaxies
in 55 nearby clusters. They find that if they normalize the clustercentric distances by a characteristic
cluster radius $R_c^{opt}$, corresponding to the radius within which the density of
bright galaxies has some fixed value, the morphology-radius relation is ``universal''
and  does not vary with                  
cluster richness, velocity dispersion or X-ray luminosity. The fraction of spiral
and irregular galaxies increases from near zero at the cluster center to $\sim 0.6$ at the
outskirts of the cluster. 

Trends in morphology have also been studied in a number of high redshift clusters 
( e.g. Abraham et al 1996;
Dressler et al 1997 ; Balogh et al 1998; Couch et al 1998;  Van Dokkum et al 2000).
 
An increase in the fraction of
early-type galaxies towards the centres of clusters is observed in all cases, but
the detailed evolution of the morphology-radius relation to high redshift remains a
subject of considerable controversy.
Dressler et al. (1997)  claim that the fraction of elliptical galaxies in rich clusters
remains constant with redshift, but the fraction of S0 galaxies at $z \sim 0.4$ 
is a factor 2-3 times smaller
than at present with a proportional increase in the numbers of spirals.
This has recently been confirmed by a  separate study by Fasano et al (2000).
Van Dokkum et al (2000) dispense with the distiction between elliptical and S0, and study the evolution
of the fraction of early-type galaxies in rich clusters. They claim that there is a clear trend for
high-redshift clusters to have lower early-type fractions than low-redshift clusters.
Within a fixed physical radius of 300 $h^{-1}$ kpc, the early-type fraction appears to decline
by a factor of two from $\sim 0.8$ at $z=0$ to $\sim 0.4$ at $z=0.8$.

One major difficulty in comparing our morphology gradients                                  
with observational data is that it is not clear whether the classification of galaxies   
according to $B/T$ corresponds in any simple way to ``by eye'' classifications 
according to Hubble type.  
In this section, we compare our $z=0.2$ morphology-radius relation with the CNOC1 sample 
of cluster galaxies for which classifications according to $B/T$ are
available for $\sim$75\% of the sample (the galaxies that are exluded show no
correlation with redshift, emission line properties or colour (Balogh et al 1998)).
We select only galaxies with R-band absolute magnitudes brighter than -20.5 in both
the data and in the simulations. Cluster membership is defined as described
previously.

\begin{figure}
\centerline{
\epsfxsize=7cm \epsfbox{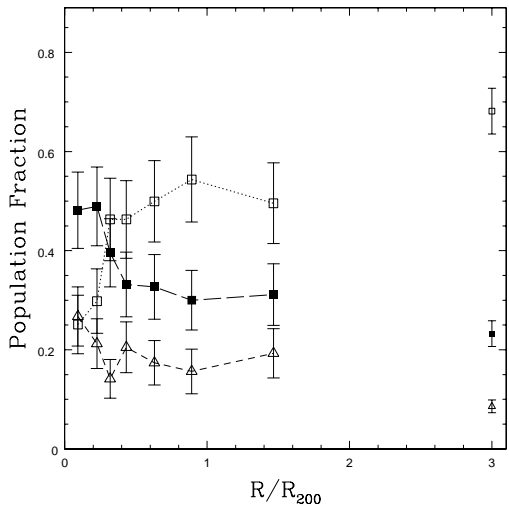}
}
\caption{\label{morphs1}
The fraction of bulge-dominated (solid squares), disk-dominated (open squares) 
and intermediate-type (open
triangles) galaxies is plotted as a function of scaled projected radius for the CNOC1 clusters.
The error bars present the  Poisson errors.
The symbols on the right-hand side of the plot indicate the fractions in the field.}
\small
\end {figure}

\begin{figure}
\centerline{
\epsfxsize=7cm \epsfbox{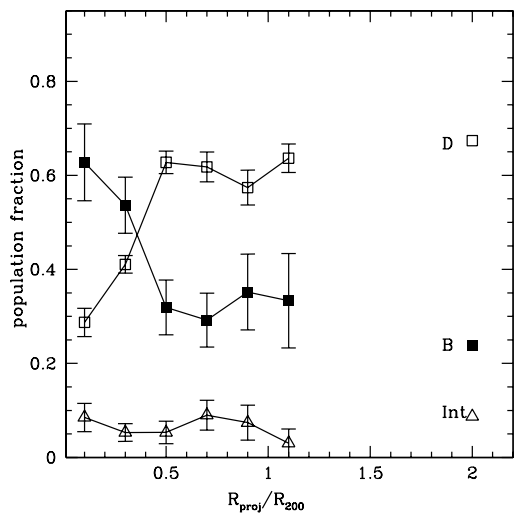}
}
\caption{\label{balmorph}
As in figure \ref {morphs1}, except for the simulated clusters. The boundaries between the morphological
classes were chosen so that the field fractions match those in figure \ref {morphs1}.}
\small
\end {figure}

The fractions of galaxies in each class as a function of projected clustercentric radius 
are shown in figures \ref{morphs1} and \ref{balmorph}. The gradient of bulge-dominated
galaxies agrees remarkably well with the observations. It exhibits a steep drop from the cluster centre
out to a radius of $ \sim 0.5 R_{200}$ and then flattens in the outer
regions of the cluster.  The fractions of disk-dominated and intermediate
galaxies do not agree as well. There are too many disk-dominated galaxies in the cluster
and not enough intermediate systems, suggesting that additional processes may lead to 
bulge formation and/or disk destruction in the cluster.

\normalsize

\normalsize

\begin{figure}
\centerline{
\epsfxsize=7cm \epsfbox{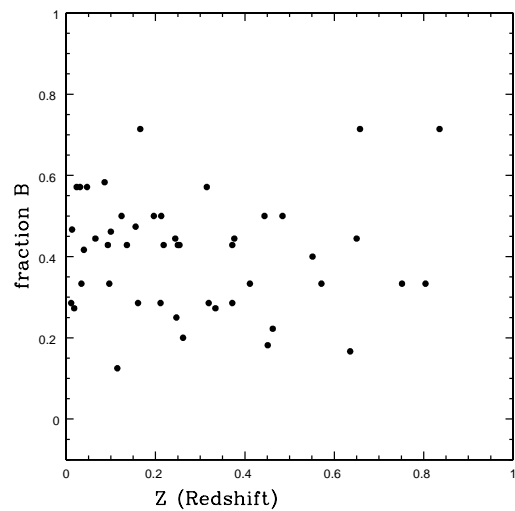}
}
\caption{\label{scatmorph}
\small
The fraction of galaxies inside $R_{200}$ with $B/T > 0.4$ is plotted against redshift for
each cluster in the sample.}
\end {figure}

The morphology-radius relation in the simulation does not evolve noticeably with
redshift. This is shown again in fig \ref{scatmorph}, where we plot the  fraction of
bulge-dominated ($B/T> 0.4$) galaxies inside $R_{200}$ for each cluster as a function of its redshift.
There is a fairly large cluster-to-cluster scatter,
but no evidence of a decline in the fraction of bulge-dominated galaxies at high redshifts.
However, as we show in  section 8.2, the fraction of bulge-dominated galaxies depends on cluster
mass, so this result will be sensitive to how clusters are selected observationally.               

\normalsize

\subsection {Distribution of morphologies as a function of cluster mass}

\begin{figure}
\centerline{
\epsfxsize=8cm \epsfbox{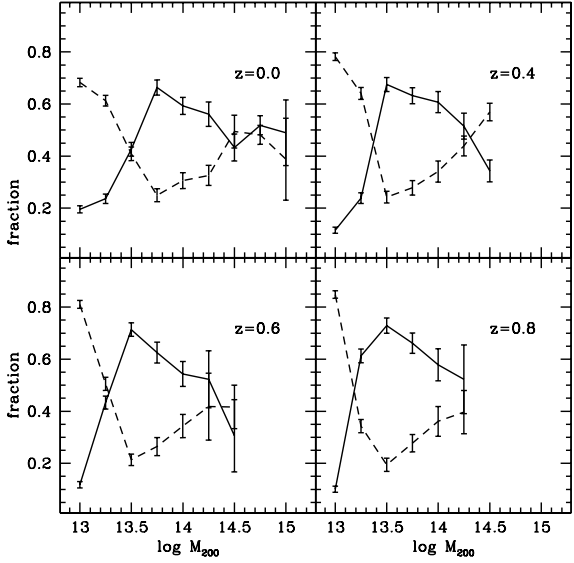}
}
\caption{\label{morphmass}
The solid line shows the fraction of galaxies with $M_R < -20.5$ 
and $B/T>0.4$ located inside $R_{200}$ as a function of
the virial mass of the cluster. The dashed line is for galaxies with $B/T < 0.2$.
Error bars show the $1-\sigma$ scatter between different clusters of the same mass
in the simulation.}
\small
\end {figure}

In figure \ref{morphmass} we show how the fractions of bulge-dominated (B)
and disk-dominated (D)  galaxies vary
as a function of the mass of the cluster. We have selected galaxies with $M_R < -20.5$
located  less than $R_{200}$
from the cluster centre.

The B fraction shows a pronounced peak for clusters  
$ \sim 3 \times 10^{14} M_{\odot}$ {\em and then declines} for masses larger than this.
The D fractions mirror this trend, showing a significant increase for the most
massive clusters.

Whitmore, Gilmore \& Jones (1993),
find no dependence of galaxy morphology on the X-ray luminosity or the
velocity dispersion of the clusters in their sample. Figure \ref{morphmass} indicates 
that the dependence of morphology on cluster mass is stronger
at high redshift than at low redshift, so it would be very interesting to repeat their analysis
for a sample of higher redshift clusters. Fasano et al (2000) find a strong dependence of the
relative numbers of elliptical and S0 galaxies on cluster type. In particular, clusters in which the
population of ellipticals is strongly concentrated have fewer S0 galaxies.

\normalsize

\subsection {The relation between morphology and star formation rate } 
\begin{figure}
\centerline{
\epsfxsize=8cm \epsfbox{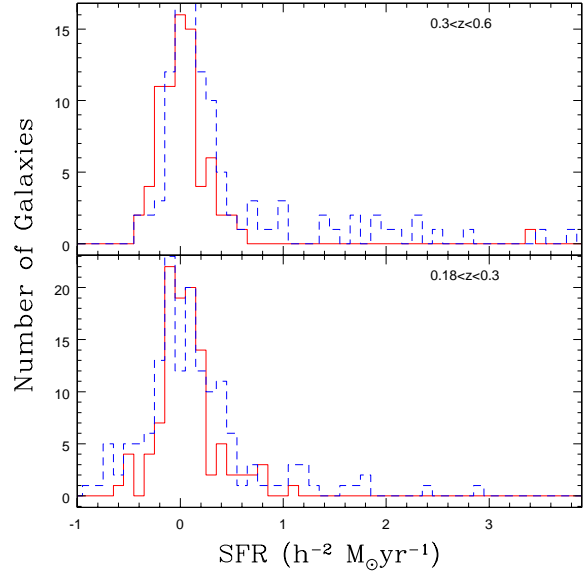}
}
\caption{\label{sfr_morph}
The solid line shows the star formation rate distribution of bulge-dominated
galaxies in the CNOC1 clusters, while the dashed line shows the star formation rate
distribution of disk-dominated systems.  
We have selected galaxies with R-band magnitudes less than -20.5.}
\small
\end {figure}

\begin{figure}
\centerline{
\epsfxsize=8cm \epsfbox{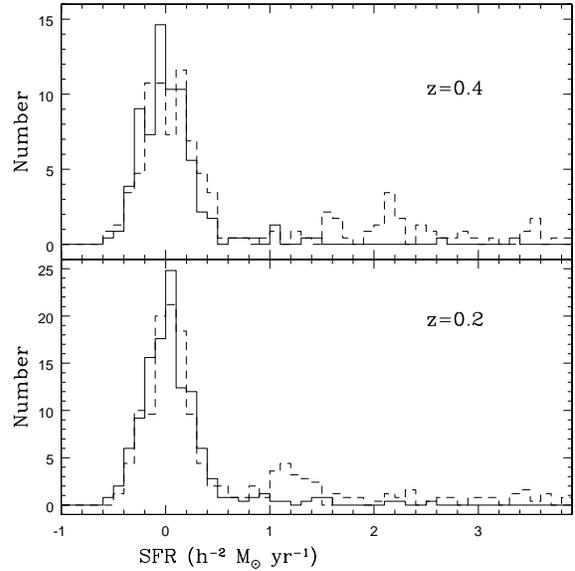}
}
\caption{\label{balsfmorph}
As in figure \ref {sfr_morph}, except for the simulated clusters.}
\small
\end {figure}

In previous sections we have compared the star formation rate and morphology gradients
in our simulated clusters with the CNOC1 data and have found reasonably good agreement. 
Here we compare the star formation rate distributions of disk-dominated (D) and bulge-dominated (B)
galaxies in clusters.
Figures \ref{sfr_morph} and \ref{balsfmorph} compare the star formation
rate distributions of galaxies in these these two classes at two different redshifts.  
Unlike bulge-dominated galaxies, the SFR distribution of disk galaxies exhibit a pronouced tail of higher
star formation rate systems. Moreover, the fraction of disk galaxies in the tail appears
to increase to higher redshift.  
This increase is seen in both the
simulated clusters and in the CNOC1 clusters, but appears somewhat stronger in the latter.

\normalsize

\normalsize

\section {The kinematics of cluster galaxies}

\begin{figure}
\plotfiddle{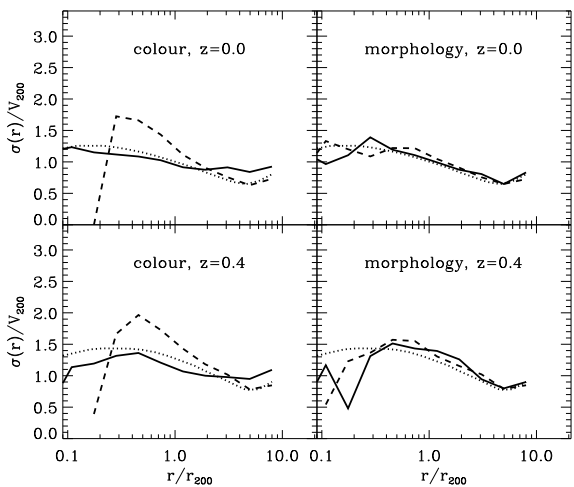}
           {0.3\vsize}              
           {90}                
           {40}                 
           {40}                 
           {150}               
           {-10}                
\caption{Evolution of the differential velocity dispersion profiles for clusters
with $M_{200}>10^{14}M_\odot$. Solid lines are the dark matter profiles;
dashed (dotted) lines are the profiles of red (blue) galaxies 
or bulge-dominated ( disk-dominated) galaxies. In the left panels, the galaxy sample is separated according
to the median rest-frame colour $B-V$ at each redshift; in the right
panels, B (D) type galaxies
have bulge-to-total luminosity ratio $B/T>0.4$ ($B/T<0.2$).
The galaxy sample only includes galaxies brighter than $M_R=-20.5$. 
Note that both red and bulge-dominated galaxies track the differential velocity dispersion
profiles of the dark matter component reasonably well both at low and high redshifts.}
\label{sigma}
\end{figure}

The existence of combined information on the spatial and kinematic distributions  
of cluster galaxies as a function of colour and morphological type  
provides fundamental constraints on how clusters were formed  
and on how their galaxy populations were accreted 
(e.g. Dickens \& Moss 1976; Colless \& Dunn 1996; Mohr et al. 1996; Carlberg et al 1997;
Biviano et al 1997; Fisher et al 1998;  De Theije \& Katgert 1999).  
The line-of-sight velocity dispersions of star-forming galaxies 
are $\sim 20 -40 \%$ larger than the velocity dispersions of 
non-star forming galaxies. This result suggests that the star-forming  
galaxies are on fairly radial, `first-approach' orbits towards the central regions of
their clusters, and are not yet in equilibrium with the population of non-star forming
galaxies.
Figure \ref{sigma} 
shows the differential velocity dispersion profiles for red/blue and bulge-dominated/disk-dominated
galaxies in our simulated clusters at two different redshifts.
Velocities are in units of $V_{200}$, the cluster circular velocity at $R_{200}$.

As discussed previously, blue galaxies have just been accreted by the cluster. They are 
less bound and their velocity dispersions in the central regions of clusters are
$\sim 30\%$ larger than those of red galaxies.
Figure \ref{sigma} indicates that there  is rather little change in the velocity
profiles of galaxies between low and high redshift clusters if we split
our sample at the median colour at each redshift.

In contrast, there is almost no difference in the differential velocity dispersion profiles of 
bulge-dominated and disk-dominated cluster galaxies. In our models, the morphological evolution of cluster
galaxies and their star formation histories are decoupled. 
We have shown that clusters contain a substantial population of red disk-dominated galaxies
that were accreted more than $\sim 1-2$ Gyr ago and are thus in dynamical equilibrium
within the cluster.     
This compensates for the larger velocities of the 
recently-accreted, star-forming disk galaxies. 

\begin{figure}
\plotfiddle{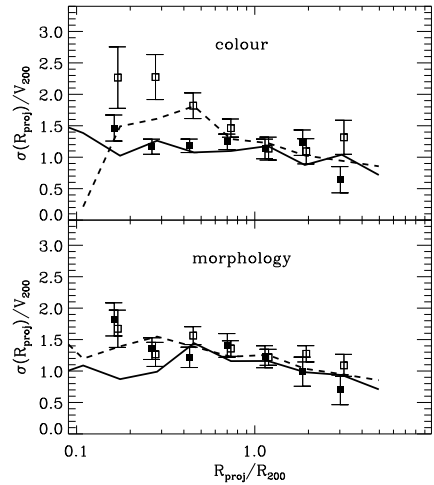}
           {0.6\vsize}              
           {90}                
           {40}                 
           {40}                 
           {150}               
           {-10}                
\caption{Differential velocity dispersion profiles of galaxies 
brighter than $M_R=-20.5$  in the CNOC1 clusters.
The top panel shows the profiles of blue (open squares) and red (filled squares) galaxies;
the rest-frame colour threshold is $B-V=0.85$. The dashed 
and the solid lines show the corresponding 
profiles of the  blue and red galaxies in the simulated clusters.
The bottom panel shows the profile of D galaxies (open squares) and B galaxies   
(filled squares) galaxies in the data. The dashed
and the solid lines show the profiles of the model D and B galaxies respectively.  
In both panels, simulated clusters are at $z=0.2$, and
error bars on the data are the standard deviations computed with the bootstrap method.}
\label{velcnoc}
\end{figure}

These results are in good agreement the CNOC1 data.                                     
Figure \ref{velcnoc} shows the differential velocity dispersions of CNOC1 cluster galaxies:  
the red and blue subsamples have clearly distinct profiles in the
central regions, whereas the bulge-dominated and disk-dominated galaxies have indistinguishable
profiles. The lines show the model predictions for model galaxies with the same selection criteria
as in the data. 

\begin{figure}
\plotfiddle{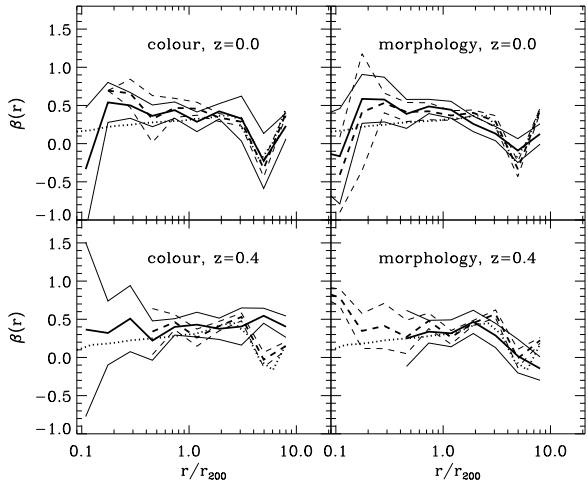}
           {0.56\vsize}              
           {90}                
           {40}                 
           {40}                 
           {150}               
           {-15}                
\caption{Evolution of the average profiles of the velocity anisotropy parameter
$\beta(r)$ in clusters
with $M_{200}>10^{14}M_\odot$. The thick lines are as for Figure \ref{sigma}.
The thin lines indicate the $1-\sigma$ bootstrap fluctuations. For clarity we omitted
the profiles where poor statistics yields a 1-$\sigma$ bootstrap range larger than 2.8.}
\label{beta}
\end{figure}

In the simulations, red and blue  galaxies do not show any 
difference in their orbital parameters: Figure \ref{beta} shows 
the profile of the velocity anisotropy parameter $\beta(r)=1-\langle v_t^2\rangle/2\langle v_r^2\rangle$,
where $v_r$ and $v_t$ are the radial and tangential components of the galaxy
velocities respectively. Both red and blue galaxies have $\beta(r)$ similar
to that of the dark matter and smaller than 0.5. This result indicates that
the velocity distribution is close to isotropic as also inferred 
for the CNOC1 clusters by  Van der Marel et al. (1999). Note that Figure \ref{beta} shows
the profile {\it averaged} over all the massive clusters in our simulation box.
Individual clusters may exhibit larger velocity anisotropies; however, we never find any 
substantial differences between blue and red galaxies. Splitting the 
galaxy sample according to the star formation rate does not 
change our conclusion.

\section {Summary and Discussion}
In this paper, we demonstrate how combining semi-analytic modelling of galaxies with
N-body simulations of cluster formation allows us to 
study spatial variations in  the colours, star formation
rates and morphologies of cluster galaxies and their evolution with redshift.          
We have shown that gradients in galaxy properties arise naturally in hierarchical models,
because mixing is incomplete during cluster assembly. 
The positions of galaxies
within the cluster are correlated with the epoch at which they were accreted.
As a result, galaxies in the cores of clusters  have lower star formation rates,
redder colours and larger bulge-to-disk ratios than galaxies in the outer regions.
We have also demonstrated that star-forming cluster galaxies have larger   
velocity dispersions than non-starforming galaxies.
Our models predict that the mean colours and star formation rates of cluster galaxies
become equal to the field values at distances of $\sim 2-3 R_{200}$ from the cluster
centre.   
 
We have compared our derived gradients with recent observational data from the CNOC1 cluster
survey. Our star formation rate and colour gradients  agree reasonably well
with the data. In agreement with  Balogh, Navarro \& Morris (2000), we find that the CNOC1 results 
are consistent with a picture in which star formation is gradually terminated over a period
of 1-2 Gyr after galaxies fall into the cluster. 
We also study the velocity dispersion profiles of cluster galaxies in our simulations
as a function of colour and find that they match the data. 

Our models are also reasonably successful in explaining the observed trends in 
galaxy morphology as a function of clustercentric radius.
For simplicity we have assumed that a galaxy's morphology is determined solely by its
history of major mergers. A major merger leads to the formation of a bulge
and any gas that cools thereafter forms a new disk. 
We have shown that this model works very well for bulge-dominated galaxies, but is
less successful in explaining the observed fractions of galaxies with intermediate
and low bulge-to-disk ratios.
In order to bring our results into agreement
with observations, some additional process must either destroy existing disks or form new     
bulges in cluster galaxies.
This suggests that ram-pressure stripping or galaxy harassment may affect the morphologies
of galaxies, but have little effect on their star formation rates.
 
We have also studied how star-formation rates of cluster galaxies vary as a function of bulge-to-disk
ratio and as a function of redshift in the CNOC1 and simulated
clusters. The star formation rate distributions of 
bulge-dominated and disk-dominated cluster galaxies are peaked near zero, 
but disk galaxies exhibit a tail
of higher star formation rate systems. The fraction of the population in this tail
appears to increase to higher redshift. The star formation rates of bulge-dominated
galaxies evolve much more weakly with redshift.

In our models, the evolution of the star formation rates and the morphologies of cluster galaxies
are largely decoupled. 
We predict that the colours of cluster galaxies are largely independent of the mass of
the cluster, but that there is a strong dependence of morphology on cluster mass. In particular,
more massive clusters are predicted to contain a smaller fraction of bulge-dominated
galaxies formed by major mergers.

We find that the  major disagreement between the simulations and the CNOC1 data occurs not
for cluster galaxies, but for field galaxies.
Bright CNOC1 field galaxies evolve dramatically in star formation rate and in colour
over the redshift range $0.18 < z < 0.55$. This is not seen in the simulations. 
We have argued that
this rapid evolution in star formation rate in the field may also be manifested as a 
strong apparent increase in the number of blue galaxies in clusters at high redshift, because a large
fraction of the blue cluster members in our simulation are in fact interlopers from
the field.
We have chosen not to delve further into the issue of field galaxy evolution in this paper,
because the CNOC1 field sample is rather small and is not selected in a completely unbiased way due
to the fact that the fields always contain a rich cluster.
It is therefore not the ideal sample for a detailed comparison between theory and observations.

Finally, our analysis demonstrates how important it is for simulation data to be analyzed
the same way as the observations. The  star formation and morphology gradients
plotted as a function of projected radius can differ substantially from the true physical gradients. 
Although interlopers from the field may be a small fraction
of the total cluster sample, they may be a more significant component of the blue population,
which is intrinsically rare in the centres of clusters.

\vspace{0.8cm}

\large
{\bf Acknowledgments}\\
\normalsize
The simulations in this paper were carried out using codes kindly made available by the Virgo Consortium.
We especially thank J\"org Colberg and  Adrian  Jenkins for help in carrying them out.
GK thanks Pieter van Dokkum and Richard Ellis for helpful discussions.
We would also like to thank members of the CNOC1 team for allowing us to use their data
in advance of full publication. MLB gratefully acknowledges support from a PPARC rolling
grant for extragalactic astronomy and cosmology at the University of Durham.

\end {document}